\documentclass
{amsart}
\usepackage{amssymb,amsmath,graphicx}
\theoremstyle{definition}

\theoremstyle{remark}

\numberwithin{equation}{section}

\begin{document}
\title [SUGRA Interactions]{SUGRA Interactions within Flavor Triplets}
\author{J. Towe}
\address{Department of Physics, The Antelope Valley College, Lancaster, CA 93536}%
\email{jtowe@avc.edu}\
\begin{abstract}
A specific new quark permits that flavor generations constitute a
representation of the 3-dimensional SU(3) symmetry that
characterizes the $Z_{3}$ orbifold. In this context, color and
local supersymmetry bind triplets and 4-tuplets into composite
fields of spin 3/2 and spin 2; and the symmetry $E_{8}$ that
characterizes (the observable sector of) 10-spacetime is
interpreted as having reduced to SU(5)XSU(3)$_{3-D}$; e.g. to a
locally supersymmetric unification of flavor generations and of
quark colors and basic $I_{3}$ classes that are devoid of color
and hypercharge. In this context superunified interactions occur
to color bound quarks that are experiencing asymptotic freedom
within baryons. Quark-lepton transitions are produced, but quickly
reverse, preserving flavor triplets. The symmetry consisting of
six quark classes and six lepton classes is also maintained
because the predicted quark is an anomalous (left-handed) version
of the strange quark.
\end{abstract} \maketitle
$ $\\[-06pt]
\section {Introduction}\label{S:intro}
The current status of high energy physics is plagued by a
difficult problem: that SUSY SU(5) predicts proton decay that has
not, after several years of intensive experimental activity, been
observed. This problem may occur because local supersymmetry is
relevant at energy levels lower than traditionally assumed.
Specifically, there are indications that supergravity is
significant at the GUT level--around $10^{16}$ GeV. Thus it may be
that SUSY GUTs should be replaced by, or modified as aspects of
supergravity. However, the only known finite supergravity theory
is that which emerges from the theory of heterotic superstrings.
Unobserved superpartners can probably be relegated to the hidden
sector of 10-spacetime, but this is a discussion for another
occassion. Assumming that this can be done, the question arises
regarding a physical representation of the 3-dimensional SU(3)
symmetry that characterizes the $Z_{3}$ orbifold. This
representation can be formulated if one admits a specific new
quark: an anolalous, left-handed version of the strange
quark--anomalous because this quark associates with a strangeness
number of zero (left-handedness implies that $I_{3}\neq0$, which,
in the context of the postulated symmetry requires that Y=1/3.
Thus, Y=B+S implies that 1/3=1/3+S, or that S=0). The postulated
quark is characterized as a strange quark because associated
quantum numbers, other than strangeness, are those of the strange.
In the proposed model moreover, the new quark is the generational
partner of the charmed quark. The newly introduced quark permits
that quark generations be organized into three
triplet--anti-triplet configurations, all of which contain the
right-handed strange and left-handed anti-strange quarks [J. Towe,
1997]. 

\par
If the above described triplets are symmetrically combined, so that
none is distinguished from any other, and so that degeneracy is
avoided (so that the strange and anti-strange quarks occur only
once), then one obtains a physical representation of a
3-dimensional SU(3) symmetry, which is spanned by the basis:
\begin{equation} \label{E:int}
e^{in2\pi/3}I_{3}, Y: n = 1, 2, 3.
\end{equation}
This configuration is depicted by the first figure.
(The proposed model should be compared with the traditional theory
of quark generations [D. Nordstrom, 1992].)
\par
If the symmetry that is spanned by expression 1.1 is imbedded into
the point group of the relevant torus, then this 3-dimensional
SU(3) symmetry becomes a physical representation of the $Z_{3}$
orbifold. However, the proposed model immediately elicits an
important question: Does the postulated physical representation of
$SU(3)_{3-D}$ also satisfy local supersymmetry, containing the
spin-(3/2) and spin-2 fields that are necessary for supergravity
interactions? The answer is that it can. The triplet--anti-triplet
configurations that are combined to yield the configuration
depicted by the first figure are usually regarded as of spin 1/2,
but they can be of spin 3/2 if the strange and anti-strange quarks
are anti-aligned with the other quarks and anti-quarks. This
characteristic will be imposed upon the representation that is
depicted by the first figure; i.e. upon the generic representation
of 3-dimensional SU(3) that was proposed.

In this context,
one consults the Figure 4 configuration, observing that there are
three orientations of the $I_{3}$-axis about the hypercharge axis for
which a triplet and anti-triplet lie on the ($I_{3}$-Y)-plane. One of
these orientations corresponds to the up down and strange quarks.
A second corresponds to the charmed, 7 and strange quarks (7 referring to the
predicted quark) and the third corresponds to the top bottom and strange
quarks.

\par
For each of the above described orientations of the $I_{3}$ axis,
there are 4 quarks and 4 anti-quarks that lie off the
($I_{3}$-Y)-plane; and two options for organizing these into
spin-2 4-tuplets and anti-4-tuplets; e.g. for the UDS triplet,
there are the 4-tuplet G=C7TB and anti-4-tuplet
$\overline{G}$=$\overline{C}\overline{7}\overline{T}\overline{B}$;
and the 4-tuplet g=C$\overline{7}\overline{T}B$ and anti-4-tuplet
$\overline{g}$=$\overline{C}$7T$\overline{B}$. Note that G and
$\overline{G}$ are respectively characterized by a charge of 2/3 and
one color, and by a charge of -2/3 and one anti-color; and that g
and $\overline{g}$ are of charge zero and colorless.
Thus, there is a total of 6 spin-2 4-tuplets. Because the adjoint
representation of the GL(4) group contains 6 elements, it is postulated
that the 6 proposed spin-2 4-tuplets constitute an irreducible, adjoint
representation of GL(4), just as the spin-(1/2) and spin-(3/2) baryons
constitute irreducible representations of SU(3). The 6 proposed 4-tuplets
are therefore regarded as composite, spin-2 fields. G-fields will
subsequently be referred to a type I fields and g-fields as type
II fields. One can now answer the question that was posed above: The proposed version of the Figure 4 flavor representation
of 3-dimensional SU(3), in which triplets are of spin-(3/2) does satisfy
local supersymmetry, containing the spin-2 and spin-(3/2) fields that
are necessary for supergravity interactions.
\par
In the above postulated context the symmetry $E_{8}$,
corresponding to the observable sector of 10-spacetime, is
interpreted as having reduced to SU(5)XSU(3)$_{3-D}$; i.e. is
interpreted as a unification of the fermionic flavor generations with
quark colors and the basic $I_{3}$ classes that are devoid of
hypercharge and color.

\par
The nature of basic (first order) supergravity interactions is
determined by the locally supersymmetric Lagrangian, which prescribes
that SUSY vertices be generated by action of a graviton vertex
operator upon a gravitino (spin-(3/2) field), which produces a
graviton (spin-2 field). The specific interactions that will be
considered here are 2nd order corrections of the basic, locally
supersymmetric interactions that are required by the locally
supersymmetric Lagrangian; and are based upon the asymptotic freedom
that is experienced by color-bound quarks at close range. Specifically,
it is proposed that interactions between the proposed spin-2 fields
and spin-(3/2) triplets are actually interactions between spin-2
fields and 'valance quarks' (quarks that experience asymptotic freedom
within triplets, and in this context, absorb or radiate
spin-2 fields or anti-fields). If the spin-2 and spin-(3/2) fields that
we have proposed enter into this kind of interaction, then quark-lepton transitions
can occur, while baryon decay is avoided. The proposed interactions will now
be considered in some detail.

\section{Supergravity Interactions}\label{S:intro}
The action of pure supergravity is given by a sum of two
integrals:
\begin{equation} \label{E:int}
\frac{-1}{2\kappa}\int d^{4}x|det e|R-\frac{1}{2}\end{equation}
(e=$e_{\mu}^{m}$ is the verbein, where $\mu$ is a world sheet
index and m is a local Lorentz index) and
\begin{equation} \int d^{4}x \epsilon^{\mu\nu\rho\sigma}\overline{\psi}_{\
mu}\gamma_{5}\overline{D}_{\rho}\psi_{\sigma}
\end{equation}
i.e. by the sum of the standard Einstein action where R represents
the curvature scaler, and the action of the Rarita-Schwinger
field, which is covariantized in terms of a covariant derivative
to which an extra term has been added. This covariant derivative
differs from the ordinary covariant derivative by a term quadratic
in the Rarita-Schwinger field, which is necessary to achieve
invariance of the action at order $\kappa^{2}$. The action is
invariant, to this order of $\kappa$, under local supersymmetry
transformations $e^{i\epsilon}(x)$Q, where $\epsilon$ is a Majorana spinor parameter that is a
function of spacetime position, and Q is a Majorana generator of
supersymmetry.
\par

The interaction vertex that arises from this action involves a
gravitino state of momentum k that corresponds to a vector-spinor
$u^{\mu}$. In the context of heterotic string theory, graviton
emission from this gravitino ground state is produced by
interaction of the gravitino with a bosonic right-moving
(Neveu-Schwarz) prescription, which is tensored with a fermionic
left-moving (Ramond) prescription. This 'graviton vertex operator'
is given by a sum of the following terms:
\begin {equation} \label{E:int}
\epsilon_{\mu\nu}[\partial_{\tau}X^{\mu}_{R}(0)]\end{equation} and
\begin{equation} \frac{1}{2}[\psi^{\mu}_{R}k
\psi_{R}(0)]\psi^{\nu}_{L}(0)e^{-ikX}
\end{equation}
[J. Bailen, 1994].
\par
As explained above, the interactions to be considered here are 2nd
order corrections of the basic locally supersymmetric interactions
(those consisting of the vertices that are described above) which
corrections are due to the asymptotic freedom experienced at short
range by quarks within flavor triplets. In practice, each
interaction of interest involves an asymptotically free quark that
simultaneously absorbs a $\overline{G}$-field (a type I anti-field
of spin-2) and radiates a $\overline{g}$-field (a type II
anti-field of spin-2). Such interactions always produce leptons
[J. Towe, 2003]. There are three options for the occurrence of
this fundamental interaction. One is that represented when
$\overline{G}_{R}$ is absorbed and $\overline{g}_{R}$ is radiated
by a down quark, $D_{L}$, producing an LH electron.
A second option is provided by the case in which
$\overline{G}_{R}$ is absorbed, and $\overline{g}_{R}$ is
simultaneously radiated by a $U_{L}$, to produce a colorless, LH,
spin-(1/2) particle with a charge of zero; i.e. an LH electron's
neutrino. 
A third realization of the
postulated interaction is provided by the case in which a
$\overline{G}_{R}$ is absorbed, and a $\overline{g}_{R}$ is
simultaneously radiated by a strange quark within an $\Omega$
(triplets SSS), producing a right-handed electron. 
The $\overline{G}$ and $\overline{g}$
fields are anti-aligned with the valance quark in 
the interaction just described, so that they behave essentially as
LH fields. 
\par
The above described interactions 
are idealized in that supersymmetric interactions occur at
vertices of triplet separation or recombination. In these special
cases, the interactions can be mediated exclusively by spin-2
fields; but if a supersymmetric interaction occurs between a
vertex of triplet separation and recombination, then a spin-1
field (an element of the adjoint representation of SU(5)) must be
absorbed together with a type I spin-2 anti-field and a spin-1
field must be radiated together with a type II, spin-2 anti-field.
Finally, if a baryon is of spin-(1/2), then a spin-1 field must be
absorbed and radiated by the valance quark even if the
supersymmetric interaction occurs at a vertex of triplet
separation or recombination.




Specifically, inspecting the second figure, one observes that the
admission of an X-particle permits that vertices preserve
supersymmetry. In general then, GUT interactions are necessary to
permit the general implementation of the proposed
superunification, but they are only admitted as aspects of locally
supersymmetric interactions.
\par
\section{Conclusion}\label{S:intro}
It was shown that if a specific new quark is introduced, and if
the superpartners of the fermions are relegated to the hidden
sector of the ($E_{8}XE_{8}$) 10-spacetime domain, then the quark
generations (in the observable sector) can be organized into three
triplet--anti-triplet configurations that contain the strange and
anti-strange quarks as common elements. Secondly it was argued
that if these triplet--anti-triplet configurations are
symmetrically combined, so that none is distinguished from any
other (and the degeneracy disappears: strange and anti-strange
quarks occurring only once), then they form a realization of the
relevant 3-dimensional SU(3) symmetry. Thirdly it was observed
that the proposed flavor configuration also satisfies local
supersymmetry, containing the necessary spin-(3/2) and spin-2
fields.


\par
In the above described context, the symmetry $E_{8}$ of (the
observable sector of) 10-spacetime, was interpreted as having reduced
to SU(5)X$SU(3)_{3-D}$; i.e. was interpreted as a unification of the
fermionic flavor generations with quark colors and basic $I_{3}$
classes that are devoid of color and hypercharge.
\par
It was argued that interactions of spin-2 fields with baryons are
actually 2nd order corrections of basic supergravity interactions--
corrections in which spin-2 fields are absorbed and radiated by
valance quarks (quarks that experience asymptotic freedom within triplets
and, in this context, undergo gravitational interactions). Specifically,
it was argued that asymptotic freedom permits the coupling of a U, D or S
singlet to a spin-2 field 
It was emphasized that these
couplings result in quark-lepton transitions that quickly reverse, preserving
quark triplets. It was observed that the interactions depicted by figures
5-7 are idealized (probably naive) in that supersymmetric interactions occur
at vertices of triplet separation or recombination. It was emphasized that if
a locally SUSY interaction occurs between a vertex of triplet separation and a
vertex of recombination, then that interaction must involve the absorption
of a spin-1 field together with a type I spin-2 anti-field, and the
radiation of a spin-1 field together with the radiation of a type II,
spin-2 anti-field. Finally, it was stated that if a supersymmetric
interaction occurs within a baryon of spin-(1/2), then a spin-1 field
must be involved even if the interaction does occur at a triplet
separation vertex (or recombination vertex). It was therefore concluded
that GUT interactions are necessary for the general implementation of
the proposed superunification, but such interactions were only admitted
as aspects of local supersymmetric interactions. The postulated interactions
are remarkable because they produce quark-lepton transitions while avoiding
baryon decay. Six-quark--six-lepton symmetry is also preserved because the
required quark is an anomalous (left-handed) version of the strange
quark.
\par

\end{document}